\documentclass[aps,preprint]{revtex4}%
\usepackage{amsfonts}
\usepackage{amsmath}
\usepackage{amssymb}
\usepackage{graphicx}%
\providecommand{\U}[1]{\protect\rule{.1in}{.1in}}
\newtheorem{theorem}{Theorem}

\newtheorem{remark}[theorem]{Remark}

\newenvironment{proof}[1][Proof]{\noindent\textbf{#1.} }{\ \rule{0.5em}{0.5em}}
\begin{document}
\preprint{ }
\title[ ]{Algebraic structures, physics and geometry from a Unified Field Theoretical\ framework}
\author{Diego Julio Cirilo-Lombardo}
\affiliation{Bogoliubov Laboratory of Theoretical Physics, Joint Institute for Nuclear
Research, Dubna 141980, Russian Federation}
\pacs{PACS number}

\begin{abstract}
Starting from a Unified Field Theory (UFT)\ proposed previously by the author,
the possible fermionic representations arising from the same spacetime are
considered from the algebraic and geometrical viewpoint. We specifically
demonstrate in this UFT general context that the underlying basis of the
single geometrical structure $P\left(  G,M\right)  \,$\ (the principal fiber
bundle over the real spacetime manifold $M$ with structural group G)
reflecting the symmetries of the different fields carry naturally a
biquaternionic structure instead of a complex one. This fact allows us to
analyze algebraically and to interpret physically in a straighforward way the
Majorana and Dirac representations and the relation of such structures with
the spacetime signature and non-hermitian (CP) dynamic operators. Also, from
the underlying structure of the tangent space, the existence of hidden (super)
symmetries and the possibility of supersymmetric extensions of these
UFT\ models are given showing that Rothstein's theorem is incomplete for that
description. The importance of the Clifford algebras in the description of all
symmetries, mainly the interaction of gravity with the other fields, is
briefly discussed.

\end{abstract}
\volumeyear{year}
\volumenumber{number}
\issuenumber{number}
\eid{identifier}
\date[Date text]{date}
\received[Received text]{date}

\revised[Revised text]{date}

\accepted[Accepted text]{date}

\published[Published text]{date}

\startpage{101}
\endpage{102}
\maketitle
\tableofcontents

\section{Fermionic symmetry and matter fields}

Reviewing some concepts from earlier references [2], in [1] it was discussed
that according to Wigner, from the quantum viewpoint a matter field can be
defined by a spinor field $\Psi^{k}\left(  x^{\lambda}\right)  $ where $k=1,2$
; $\lambda=1,2,3,4$ and in the case of Lorentzian metric, $x^{4}=ix^{0}.$
These fields can be taken as elements of some internal space located at
$x^{\lambda}$ of the 4 dimensional spacetime manifold. The elementary field
("particle " was used by Weyl) is defined by the following transformation
property
\begin{equation}
\Psi^{\prime k}\left(  x^{\lambda}\right)  =U_{j}^{k}\left(  x,2\right)
\Psi^{j}\left(  x^{\lambda}\right)  \tag{1}%
\end{equation}
where the $U_{j}^{k}$ is the $2\times2$ matrix representation of the unitary
group $U\left(  2,\mathbb{C}\right)  $ and is a continous function of
$x^{\lambda}.$ If the argumentation given by Weyl runs in the correct way,
strictly speaking and accordling to the analysis that follows, a
biquaternionic structure is the most adequate to derive the Dirac equation.
From the algebraic viewpoint the only generalized quaternion algebra over
$\mathbb{C}$ is the ring of $2\times2$ matrices over $\mathbb{C}$ and
moreover, the Clifford algebra of a two-dimensional space with a nondegenerate
quadratic form is central, simple and it is a generalized quaternion algebra.

From what is written above, it is necessary to fully analyze the underlying
structure of the theory (and in particular the model) presented in [3-5,7,10]
not only from the physical and geometrical viewpoint but as well as first
principles. The target is clear: to find the fundamental essence of
unification as the natural world presents us.

The organization of the article is as follows: Sections II and III are devoted
to describe the spacetime manifold: Dirac structure and the relation with
Clifford algebras as the natural language of the description[11]. In Section
IV the emerging character of the biquaternionic structure and the connection
with the Dirac equation is explicitly presented and analyzed. In Section V the
Majorana representation is introduced and discussed from the point of view of
a bi-quaternionic structure. In Sections VI, VII and VIII physical aspects are
discussed considering the relationship between the structure of the tangent
space, the signature of spacetime and the algebra $\mathbb{H}$. \ Section IX
deals to the study and description of the spacetime manifold from the point of
view of supersymmetry and the Poisson structures: the Rothstein theorem is
discussed in these context. Finally in Section X conclusions and outlook are listed.

\section{The real Dirac structure of the spacetime manifold}

The principal fiber bundle (PFB) $P\left(  G,M\right)  $ with the structural
group $G$ determines the (Dirac) geometry of the spacetime. We suppose now $G$
with the general form%
\begin{equation}
G=\left(
\begin{array}
[c]{cc}%
A & B\\
-B & A
\end{array}
\right)  ,\text{ \ \ \ }G^{+}G=I_{4}=\left(
\begin{array}
[c]{cc}%
\sigma_{0} & 0\\
0 & \sigma_{0}%
\end{array}
\right)  \tag{2}%
\end{equation}
$A,B\ $\ $2\times2$ matrices and containing a manifestly symplectic structure.
Consequently, there exists a fundamental tensor $J_{\mu}^{\ \ \lambda
}J_{\lambda}^{\ \ \nu}=\delta_{\mu}^{\nu}$ invariant under $G$ with structure
\begin{equation}
J=\left(
\begin{array}
[c]{cc}%
0 & \sigma_{0}\\
-\sigma_{0} & 0
\end{array}
\right)  \tag{3}%
\end{equation}
of such manner that%
\begin{equation}
G=\left(
\begin{array}
[c]{cc}%
A & B\\
-B & A
\end{array}
\right)  =AI_{4}+BJ\tag{4}%
\end{equation}
Where however, there exists a Lorentzian metric $g_{\lambda\mu}$ [31], i.e.,
the metric of a curved spacetime manifold with signature (+ - - -), that is
also invariant under $G$ due its general form (2). Finally, a third
fundamental tensor $\sigma_{\lambda\mu}$ is also invariant under $G$ where the
following relations between the fundamental tensors are%
\begin{equation}
J_{\lambda}^{\ \ \nu}=\sigma_{\lambda\mu}g^{\lambda\nu},\text{ \ \ \ \ \ \ }%
g_{\mu\nu}=\sigma_{\lambda\mu}J_{\nu}^{\text{ }\lambda},\text{ \ \ \ \ \ \ }%
\sigma_{\lambda\mu}=J_{\lambda}^{\text{ }\nu}g_{\mu\nu}\tag{5}%
\end{equation}
where%
\begin{equation}
g^{\lambda\nu}=\frac{\partial g}{\partial g_{\lambda\nu}}\text{ \ \ }\left(
g\equiv\det(g_{\mu\nu})\right)  \tag{6}%
\end{equation}
Then, the necessary fundamental structure is given by
\begin{equation}
G=L\left(  4\right)  \cap Sp\left(  4\right)  \cap K\left(  4\right)  \tag{7}%
\end{equation}
which leaves concurrently invariant the three fundamental forms
\begin{align}
ds^{2} &  =g_{\mu\nu}dx^{\mu}dx^{\nu}\tag{8}\\
\sigma &  =\sigma_{\lambda\mu}dx^{\lambda}\wedge dx^{\mu}\tag{9}\\
\phi &  =J_{\nu}^{\text{ }\lambda}w^{\nu}v_{\lambda}\tag{10}%
\end{align}
where $w^{\nu}$ are components of a vector $w^{\nu}\in V^{\ast}:$ the dual
vector space. In expression (5) $L\left(  4\right)  $ is the Lorentz group in
4D, $Sp\left(  4\right)  $ is the Symplectic group in 4D real vector space and
$K\left(  4\right)  $ denotes the almost complex group that leaves $\phi$ invariant.

For instance, $G$ leaves the geometric (Clifford) product invariant
\begin{align}
\gamma_{\mu}\gamma_{\nu} &  =\frac{1}{2}\left(  \gamma_{\mu}\gamma_{\nu
}-\gamma_{\nu}\gamma_{\mu}\right)  +\frac{1}{2}\left(  \gamma_{\mu}\gamma
_{\nu}+\gamma_{\nu}\gamma_{\mu}\right)  \tag{11}\\
&  =\gamma_{\mu}\cdot\gamma_{\nu}+\gamma_{\mu}\wedge\gamma_{\nu}=g_{\mu\nu
}+\sigma_{\mu\nu}\tag{12}%
\end{align}
where the $\gamma_{\mu}$ are now regarded as a set of orthonormal basis
vectors, of such a manner that any vector can be represented as $\mathbf{v}%
=v^{\lambda}\gamma_{\lambda}$ and the invariant (totally antisymmetric) tensor
as
\begin{equation}
\varepsilon_{\alpha\beta\gamma\delta}\equiv\gamma_{\alpha}\wedge\gamma_{\beta
}\wedge\gamma_{\gamma}\wedge\gamma_{\delta}\tag{13}%
\end{equation}

In resume, the fundamental structure of the spacetime is then represented by
P$\left(  G,M\right)  ,$ where $G$ is given by $(5),$ which leaves the
fundamental forms invariant $(5),$ implying that%
\begin{align}
\nabla_{\lambda}g_{\mu\nu}  &  =0\tag{14}\\
\nabla_{\nu}\sigma_{\lambda\mu}  &  =0\tag{15}\\
\nabla_{\lambda}J_{\nu}^{\text{ }\lambda}  &  =0 \tag{16}%
\end{align}
where $\nabla_{\lambda}$ denotes the covariant derivative of the $G$
connection. It is interesting to note that it is only necessary to consider
two of the above three equations: the third follows automatically. Then, we
will consider $(14),(15)$ because in some sense they represent the boson and
fermion symmetry respectively. Notice that this structure is naturally a
heterotic one carrying a $\mathbb{H}\left(  n\right)  $ representation of its own.

\begin{remark}
As will be clear later, there exists a kind of supermanifold underlying
structure in this UFT\ and also in other unified theories.
\end{remark}

\section{Interlude: Clifford algebras as natural language}

It has turned out that the Clifford algebras provide very promising tools for
description and generalization of geometry and physics [13, 14, 15], also
[30]. As it was pointed out before[15] there exist two kinds of the Clifford
algebras, orthogonal and symplectic [16]. In the orthogonal Clifford algebras,
the symmetric product of two basis vectors $\mathbf{v}=v^{\lambda}%
\gamma_{\lambda}$ is the inner product and it gives the orthogonal metric,
while the antisymmetric product gives a basis bivector. In the symplectic
Clifford algebras [11], the antisymmetric product of two basis vectors $q_{a}$
is the inner product and it gives the symplectic metric, whilst the symmetric
product gives a basis bivector. Both kinds of the Clifford algebras are
included into the expressions involving the three $G$ invariant forms.
Consequently, there exist in the model a boson $\leftrightarrow$fermion
symmetry and spacetime$\leftrightarrow$phase space. An interesting point that
we use but will not discuss in detail here, is that the generators of an
orthogonal Clifford algebra can be transformed into a basis (the Witt basis)
in which they behave as fermionic creation and annihilation operators. The
generators of a symplectic Clifford algebra behave as bosonic creation and
annihilation operators as it is well know [15]. Consecuently, both kinds of
operators can be united into a single structure so that they form a basis of a `superspace'.

\begin{remark}
This important fact allows to incorporate from the very fundamental structure
of the manifold M a consistent quantum theory with a clear geometrical meaning.
\end{remark}

\section{Dirac equation and $\mathbb{H}$ structure}

As we have considered previously [3-7,10], the G-structure must describe the
spinorial field through the appearance of the Dirac equation in the tangent
space. The physical choice for the structure of $G$ can be given by%

\begin{align}
G^{+}G  &  =\left(
\begin{array}
[c]{cc}%
A & B\\
-B & A
\end{array}
\right)  \left(
\begin{array}
[c]{cc}%
A & -B\\
B & A
\end{array}
\right)  =\left(
\begin{array}
[c]{cc}%
a_{0}\sigma_{0} & \sigma\cdot a\\
-\sigma\cdot a & a_{0}\sigma_{0}%
\end{array}
\right)  \left(
\begin{array}
[c]{cc}%
a_{0}\sigma_{0} & -\sigma\cdot a\\
\sigma\cdot a & a_{0}\sigma_{0}%
\end{array}
\right) \tag{17}\\
&  =\left(
\begin{array}
[c]{cc}%
\left(  a_{0}\sigma_{0}\right)  ^{2}+\left(  \sigma\cdot a\right)  ^{2} & 0\\
0 & \left(  a_{0}\sigma_{0}\right)  ^{2}+\left(  \sigma\cdot a\right)  ^{2}%
\end{array}
\right)  =\mathbb{I}_{4} \tag{18}%
\end{align}
where $a_{b}$ are physical quantities to be determined). Then,
\begin{equation}
\left(  a_{0}\sigma_{0}\right)  ^{2}+\left(  \sigma\cdot a\right)
^{2}=1\Rightarrow a_{0}{}^{2}+a_{1}{}^{2}+a_{2}{}^{2}+a_{3}{}^{2}=1 \tag{19}%
\end{equation}
and consequently the physical meaning of the coefficients $a$ are immediatly
determined:%
\begin{equation}
a_{0}=\frac{\widehat{p}_{0}}{m},a_{1}=i\frac{\widehat{p}_{1}}{m},a_{2}%
=i\frac{\widehat{p}_{2}}{m},a_{3}=i\frac{\widehat{p}_{3}}{m} \tag{20}%
\end{equation}
leading the relativistic relation%
\begin{equation}
\widehat{p}_{0}^{2}-\widehat{p}_{1}^{2}-\widehat{p}_{2}^{2}-\widehat{p}%
_{3}^{2}=m^{2} \tag{21}%
\end{equation}
where the introduction of the momentum operators $\widehat{p}_{\mu}$ and the
mass parameter $m$ was performed. For instance, from the explicit structure of
$G$ and the meaning of $a_{b}$ we obtain%
\begin{align}
G\mathbf{v}  &  =\mathbf{u}\tag{22}\\
G^{t}\mathbf{u}  &  =\mathbf{v} \tag{23}%
\end{align}
with $\mathbf{u}=\left(
\begin{array}
[c]{c}%
u^{0}\\
u^{1}\\
u^{2}\\
u^{3}%
\end{array}
\right)  $and $\mathbf{v=}$ $\left(
\begin{array}
[c]{c}%
v^{0}\\
v^{1}\\
v^{2}\\
v^{3}%
\end{array}
\right)  .$ Explicitly in the abstract form, we have ($h=0,1)$%
\begin{align}
\left(
\begin{array}
[c]{cc}%
A & B\\
-B & A
\end{array}
\right)  \left(
\begin{array}
[c]{c}%
u^{h}\\
u^{h+2}%
\end{array}
\right)   &  =\left(
\begin{array}
[c]{c}%
v^{h}\\
v^{h+2}%
\end{array}
\right) \tag{24}\\
\left(
\begin{array}
[c]{cc}%
A & -B\\
B & A
\end{array}
\right)  \left(
\begin{array}
[c]{c}%
v^{h}\\
v^{h+2}%
\end{array}
\right)   &  =\left(
\begin{array}
[c]{c}%
u^{h}\\
u^{h+2}%
\end{array}
\right)  \tag{25}%
\end{align}
Then, having $4D$ real vector space with $G$ as its automorphism such that
$G\subset L\left(  4\right)  $ determines the real structure of the Dirac
equation in the form%
\begin{align}
\left(  \gamma_{0}p_{0}-i\gamma\cdot\mathbf{p}\right)  \mathbf{u}  &
=m\mathbf{v}\tag{26}\\
\left(  \gamma_{0}p_{0}+i\gamma\cdot\mathbf{p}\right)  \mathbf{v}  &
=m\mathbf{u} \tag{27}%
\end{align}
with%
\begin{equation}
\gamma_{0}=\left(
\begin{array}
[c]{cc}%
\sigma_{0} & 0\\
0 & \sigma_{0}%
\end{array}
\right)  ,\ \ \ \ \ \gamma=\left(
\begin{array}
[c]{cc}%
0 & -\sigma\\
\sigma & 0
\end{array}
\right)  \tag{28}%
\end{equation}
where $\sigma$ are the Pauli matrices and $\mathbf{p}=\left(  \widehat{p}%
_{1},\widehat{p}_{2},\widehat{p}_{3}\right)  $

\subsection{Biquaternionic structure}

Considering the above, we see the possibility that, writing $\mathbf{u}$ and
$\mathbf{v}$ in the following form%
\begin{align}
\eta^{h}  &  =u^{h}+iu^{h+2}\tag{29}\\
\xi^{h}  &  =v^{h}+iv^{h+2} \tag{30}%
\end{align}
the Dirac equation becomes
\begin{equation}
Q\eta=\xi\text{ and }\overline{Q}\xi=\eta\tag{31}%
\end{equation}
where $Q$ and $\overline{Q}$ are the following elements of the field of the
biquaternions%
\begin{align}
Q  &  =a_{0}\sigma_{0}-i\sigma\cdot a=A-iB\tag{32}\\
\overline{Q}  &  =a_{0}\sigma_{0}+i\sigma\cdot a=A+iB \tag{33}%
\end{align}
where the upper bar is quaternionic conjugation

The Clifford algebra in real Minkowski space is $\mathbb{H}_{2}$ but its
complexification is $\mathbb{H}_{2}\otimes\mathbb{C}=\mathbb{C}_{4}$, which is
the Dirac algebra. One may use the differential form basis and the vee
$\left(  \vee\right)  $ product in order to derive results for the Dirac gamma
matrices which are useful in quantum field theory. It is interesting to see
that the complexification of the quaternionic structure is necessary to
incorporate in any theory of massive particles with spin 1/2 when we have
$\left(  \mathbb{C},4,(1,-1-1-1\right)  )[12-14].$

\section{Majorana representation for symmetric equation}

Despite having a real representation of the Dirac equation from the G
structure, we see that it is possible to perform a unitary transformation to G
for which the Dirac equation becomes with real coefficients and symmetric for
both: fermions and antifermions. Consequently, it will be important to know
how this transformation affects the underlying structure of the spacetime from
the quaternionic viewpoint. The explicit unitary transformation is%
\begin{equation}
U=U^{-1}=\frac{1}{\sqrt{2}}\left(
\begin{array}
[c]{cc}%
1 & \sigma_{2}\\
\sigma_{2} & -1
\end{array}
\right)  \tag{34}%
\end{equation}
and it was given by Ettore Majorana in 1937 [9]. The transformation changes
the four dimensional structure of $G,$ namely $a_{0}I_{4}+\gamma\cdot a$
($\gamma$ in the standard form [8]) to $a_{0}I_{4}+\gamma^{\prime}\cdot a$
with%
\begin{align}
\gamma_{3}^{\prime}  &  \rightarrow-i\sigma_{1}\otimes\sigma_{0}\tag{35}\\
\gamma_{2}^{\prime}  &  \rightarrow\left(
\begin{array}
[c]{cc}%
0 & -\sigma_{2}\\
\sigma_{2} & 0
\end{array}
\right) \tag{36}\\
\gamma_{1}^{\prime}  &  \rightarrow i\sigma_{3}\otimes\sigma_{0} \tag{37}%
\end{align}
and in order to be complete $\beta^{\prime}\rightarrow$ $\left(
\begin{array}
[c]{cc}%
0 & \sigma_{2}\\
\sigma_{2} & 0
\end{array}
\right)  .$ Explicitly
\begin{align}
G^{\prime}  &  \rightarrow\left(
\begin{array}
[c]{cc}%
a_{0}\sigma_{0}+i\left(  \sigma_{3}a_{1}+\sigma_{1}a_{3}\right)  & -\sigma
_{2}a_{2}\\
\sigma_{2}a_{2} & a_{0}\sigma_{0}+i\left(  \sigma_{3}a_{1}+\sigma_{1}%
a_{3}\right)
\end{array}
\right) \tag{38}\\
G^{T\prime}  &  \rightarrow\left(
\begin{array}
[c]{cc}%
a_{0}\sigma_{0}-i\left(  \sigma_{3}a_{1}+\sigma_{1}a_{3}\right)  & \sigma
_{2}a_{2}\\
-\sigma_{2}a_{2} & a_{0}\sigma_{0}-i\left(  \sigma_{3}a_{1}+\sigma_{1}%
a_{3}\right)
\end{array}
\right)  \tag{39}%
\end{align}
Notice that $G^{\prime}$ and $G^{T\prime}$ $\left(  G^{\prime}G^{T\prime
}=G^{T\prime}G^{\prime}=\mathbb{I}_{4}\right)  $are related by complex
conjugation, as expected due to the performed Majorana transformation, being
the relativistic relation of previous sections without changes.

\section{Non-compact fundamental $\mathbb{H}$-structure, G and the 2+2
spacetime}

In Ref.[28] we have presented a Majorana-Weyl representation that is given by
the 2 by 2 following operators
\begin{equation}
\sigma_{\alpha}=\left(
\begin{array}
[c]{cc}%
0 & 1\\
1 & 0
\end{array}
\right)  ,\quad\sigma_{\beta}=\left(
\begin{array}
[c]{cc}%
0 & -1\\
1 & 0
\end{array}
\right)  ,\quad\sigma_{\gamma}=\left(
\begin{array}
[c]{cc}%
1 & 0\\
0 & -1
\end{array}
\right)  , \tag{40}%
\end{equation}
where the required condition over such matrices $\sigma_{\alpha}\wedge
\,\sigma_{\beta}=\sigma_{\gamma},$ $\sigma_{\beta}\wedge\,\sigma_{\gamma
}=\sigma_{\alpha}$ and $\sigma_{\gamma}\wedge\,\sigma_{\alpha}=-\sigma_{\beta
}$, evidently holds (Lie group, with $\alpha,\beta,\gamma:$fixed indices)
given the underlying non-compact $SL(2R)$ symmetry.

As we have seen previously, the G-structure must describe the spinorial field
through the appearance of the Dirac equation in the tangent space. The
physical choice for the structure of $G$ can be given by%

\begin{align}
G^{+}G  &  =\left(
\begin{array}
[c]{cc}%
A & B\\
-B & A
\end{array}
\right)  \left(
\begin{array}
[c]{cc}%
A & -B\\
B & A
\end{array}
\right)  =\left(
\begin{array}
[c]{cc}%
a_{0}\sigma_{0} & \sigma\cdot a\\
-\sigma\cdot a & a_{0}\sigma_{0}%
\end{array}
\right)  \left(
\begin{array}
[c]{cc}%
a_{0}\sigma_{0} & -\sigma\cdot a\\
\sigma\cdot a & a_{0}\sigma_{0}%
\end{array}
\right) \tag{41}\\
&  =\left(
\begin{array}
[c]{cc}%
\left(  a_{0}\sigma_{0}\right)  ^{2}+\left(  \sigma\cdot a\right)  ^{2} & 0\\
0 & \left(  a_{0}\sigma_{0}\right)  ^{2}+\left(  \sigma\cdot a\right)  ^{2}%
\end{array}
\right)  =\mathbb{I}_{4} \tag{42}%
\end{align}
where we remind that $a_{b}$ are physical quantities. Then, only from the
G-structure and not from any extra assumption, we have as before
\begin{equation}
\left(  a_{0}\sigma_{0}\right)  ^{2}+\left(  \sigma\cdot a\right)
^{2}=1\Rightarrow a_{0}{}^{2}+a_{1}{}^{2}-a_{2}{}^{2}+a_{3}{}^{2}=1 \tag{43}%
\end{equation}
notice the change of sign of $a_{2}{}^{2}$due to the non compact substructure
introduced by $\sigma_{\beta}^{2}=(-i\sigma_{2})^{2}=-1$; consequently the
physical role of the coefficients $a$ cannot be easily identified as before.
We have here two possibilities:

i) if the definition is the same for the $a_{b}$, we have%
\begin{equation}
a_{0}=\frac{\widehat{p}_{0}}{m},a_{1}=i\frac{\widehat{p}_{1}}{m},a_{2}%
=i\frac{\widehat{p}_{2}}{m},a_{3}=i\frac{\widehat{p}_{3}}{m} \tag{44}%
\end{equation}
leading to the relativistic relation%
\begin{equation}
\widehat{p}_{0}^{2}-\widehat{p}_{1}^{2}+\widehat{p}_{2}^{2}-\widehat{p}%
_{3}^{2}=m^{2} \tag{45}%
\end{equation}
where the introduction of the momentum operators $\widehat{p}_{\mu}$ and the
mass parameter $m$ was performed. In such a case, evidently the signature of
the spacetime is $(+-+-)$

The structure of the Dirac equation has now the form%
\begin{align}
\left(  \gamma_{0}p_{0}+i\gamma_{2}\widehat{p}_{2}-i\gamma\cdot\mathbf{p}%
\right)  \mathbf{u}  &  =m\mathbf{v}\tag{46}\\
\left(  \gamma_{0}p_{0}+i\gamma_{2}\widehat{p}_{2}+i\gamma\cdot\mathbf{p}%
\right)  \mathbf{v}  &  =m\mathbf{u} \tag{47}%
\end{align}
with%
\begin{equation}
\gamma_{0}=\left(
\begin{array}
[c]{cc}%
\sigma_{0} & 0\\
0 & \sigma_{0}%
\end{array}
\right)  ,\ \ \ \ \ \gamma=\left(
\begin{array}
[c]{cc}%
0 & -\sigma\\
\sigma & 0
\end{array}
\right)  \tag{48}%
\end{equation}
where $\sigma$ are the representation given now by matrices (40) and
$\mathbf{p}=\left(  \widehat{p}_{1},\widehat{p}_{3}\right)  $

ii) if the definition for the $a_{b}$ is
\begin{equation}
a_{0}=\frac{\widehat{p}_{0}}{m},a_{1}=i\frac{\widehat{p}_{1}}{m},a_{2}%
=\frac{\widehat{p}_{2}}{m},a_{3}=i\frac{\widehat{p}_{3}}{m} \tag{49}%
\end{equation}
leading to the relativistic relation%
\begin{equation}
\widehat{p}_{0}^{2}-\widehat{p}_{1}^{2}-\widehat{p}_{2}^{2}-\widehat{p}%
_{3}^{2}=m^{2} \tag{50}%
\end{equation}
where the introduction of the momentum operators $\widehat{p}_{\mu}$ and the
mass parameter $m$ was performed. In such a case, evidently the signature of
the spacetime is conserved as $(+---)$with an evident emergent non hermiticity
of the respective dynamical operators.

The structure of the Dirac equation has now the form%
\begin{align}
\left(  \gamma_{0}p_{0}-\gamma_{2}\widehat{p}_{2}-i\gamma\cdot\mathbf{p}%
\right)  \mathbf{u}  &  =m\mathbf{v}\tag{51}\\
\left(  \gamma_{0}p_{0}+\gamma_{2}\widehat{p}_{2}+i\gamma\cdot\mathbf{p}%
\right)  \mathbf{v}  &  =m\mathbf{u}\nonumber
\end{align}
with%
\begin{equation}
\gamma_{0}=\left(
\begin{array}
[c]{cc}%
\sigma_{0} & 0\\
0 & \sigma_{0}%
\end{array}
\right)  ,\ \ \ \ \ \gamma=\left(
\begin{array}
[c]{cc}%
0 & -\sigma\\
\sigma & 0
\end{array}
\right)  \tag{52}%
\end{equation}
where $\sigma$ are the representation given now by matrices (40) and
$\mathbf{p}=\left(  \widehat{p}_{1},\widehat{p}_{3}\right)  .$

\begin{remark}
from the point of view of Unification there exists a kind of "duality" between
non-hermitian structures and spacetime signatures (this fact can be crucial to
understand what happens in high dimensional theories where exist an interplay
between "duality, spacetime signature and spinors phase transitions" as
described in [27])
\end{remark}

\section{Relation between spacetime signatures and related dynamics}

From the argumentation given before, if certainly there exists a precise
relation between the spacetime signatures, physically we have two related
dynamics. As it is well known, the Palatini variational principle determines
the connection required for the space-time symmetry as well as the field
equations. As we have shown in [3-5], if by construction any geometrical
Lagrangian or action yields the $G$-invariant conditions (namely, the
intersection of the 4-dimensional Lorentz group $L_{4},$ the symplectic
$Sp\left(  4\right)  $ and the almost complex group $K\left(  4\right)  )$, as
an immediate consequence the gravitational, Dirac and Maxwell equations arise
from a such geometrical Lagrangian $L_{g}$ as a causally connected closed
system. From the tangent space viewpoint, the self-consistency is given
by[3-7]
\begin{equation}
f_{\mu\nu}\equiv\frac{1}{2}\varepsilon_{\mu\nu\rho\sigma}\sigma^{\rho\sigma
}=\ast\sigma_{\mu\nu} \tag{53}%
\end{equation}
where $\sigma_{\nu\lambda}$ is related to the torsion by ${\displaystyle\frac
{1}{6}\left(  \partial_{\mu}\sigma_{\nu\lambda}+\partial_{\nu}\sigma
_{\lambda\mu}+\partial_{\lambda}\sigma_{\mu\nu}\right)  =T_{\ \nu\mu}^{\rho
}\sigma_{\rho\lambda}}$ and $f_{\mu\nu}$ can plays naturally the role of
electromagnetic field. As the simplest illustration, due to the fact that we
are in the tangent space, the second order version of the Dirac eq. takes the
familiar form$\mathbf{:}$
\begin{align}
\left\{  \left(  \widehat{P}_{\mu}-e\widehat{A}_{\mu}\right)  ^{2}-m^{2}%
-\frac{1}{2}\sigma^{\mu\nu}f_{\mu\nu}\right\}  u^{\lambda}  &  =0\tag{54}\\
\left\{  \left(  \widehat{P}_{\mu}-e\widehat{A}_{\mu}\right)  ^{2}%
-m^{2}+e\Sigma\cdot H-ie\alpha\cdot E\right\}  u^{\lambda}  &  =0 \tag{55}%
\end{align}
where we have introduced
\begin{equation}
\sigma_{\mu\nu}=\left(  \alpha,i\Sigma\right)  ,\text{ }f^{\mu\nu}=(-E,H)
\tag{56}%
\end{equation}
(corresponding to Galilean-type coordinates) and the fact that the momentum
$\widehat{p}=\widehat{P}_{\mu}-e\widehat{A}_{\mu}$ is generalized due to the
gauge freedom and the existence of a vector torsion $h_{\alpha}$ (see also
Appendix)that in the case of ref. [3-5,7,10] is the dual of a totally
antisymmetric torsion field $h_{\alpha}=\varepsilon_{\alpha}^{\text{ }\nu
\rho\sigma}T_{\text{ }\nu\rho\sigma}$. The torsion field appears as a
consequence of the existence in the very structure of the tangent space, of
the third fundamental tensor $\sigma_{\lambda\mu}$ $.$ From the above
"euristic" perspective we make the following remarks:

i) The equation is symmetric: for $u^{\lambda}$ and the same obviously for
$v^{\lambda}($remember that $\Psi=\mathbf{u}+i\mathbf{v)}.$

ii) Because the geometrical propierties of the tangent space (G-structure) are
translated to the fields and viceversa, physically the contraction
$\sigma^{\mu\nu}f_{\mu\nu}$ represents the interplay between spin and
electromagnetic field,

iii) In the case of 2+2 signature the "electromagnetic field" has 4 electric
components and 2 magnetic ones, and in the case with 3+1 signature the
quantity $E^{2}+H^{2}$ (e.g. "energy") can be negative due to the
non-hermitian character of the generalized momentum operators.

Here we can make some interlude with respect to the above results,
particularly item iii). Interestlingly with the point of view of symmetry
structure induced by G, we find a convergence of some isolate (from recent
references) results. Some of these consequences (enumerated below)\ of that
paper involving a (2 + 2) "by hand" signatures, can be explained due to the
existence of the $SL(2R)$ symmetry of a "hidden" (bi)quaternionic structure$:$

1) Bars from the viewpoint of 2t-physics [18] considered as a minimal model
the structure of (2+2)-physics

2) Since time ago, it was suspected, looking at some structures in string
theory, two dimensional black holes [19] and conformal field theory [20], that
the (2+2)-signature is deeply linked to the SL(2,R)-group.

3) the (2+2)-signature is conjectured as an important physical concept in a
number of physical scenarios, including the background for N = 2 strings
[21-22] (see also Refs [23]), Yang-Mills theory in Atiyah-Singer background
[25] (see also Refs. [26] for the mathematical importance of the
(2+2)-signatures), Majorana-Weyl spinor in supergravity [24]

In the next Section we will bring the conceptual and mathematical consistency
to the above issues.

\section{G-structure, spacetime and fields at $T_{p}\left(  M\right)  $}

It is well known that to every Lie algebra a local Lie group corresponds only
being the G-structure a global affair (important issue without answer till
today). Starting from the six dimensional group $SL\left(  2\mathbb{C}\right)
$ it contains
\begin{align}
\sigma_{1}  &  =\frac{1}{2}\left(
\begin{array}
[c]{cc}%
0 & i\\
i & 0
\end{array}
\right)  ,\text{ }\sigma_{2}=\frac{1}{2}\left(
\begin{array}
[c]{cc}%
0 & 1\\
-1 & 0
\end{array}
\right)  ,\text{ }\sigma_{3}=\frac{1}{2}\left(
\begin{array}
[c]{cc}%
-i & 0\\
0 & i
\end{array}
\right) \tag{57}\\
\rho_{1}  &  =\frac{1}{2}\left(
\begin{array}
[c]{cc}%
0 & 1\\
1 & 0
\end{array}
\right)  ,\text{ }\rho_{2}=\frac{1}{2}\left(
\begin{array}
[c]{cc}%
0 & -i\\
i & 0
\end{array}
\right)  ,\text{ }\rho_{3}=\frac{1}{2}\left(
\begin{array}
[c]{cc}%
-1 & 0\\
0 & 1
\end{array}
\right)  \tag{58}%
\end{align}

The bispinor can be constructed on the tangent space $T_{p}\left(  M\right)  $
by complexification%

\begin{equation}
\Psi^{\prime B}=U_{A}^{B}\left(  P\right)  \Psi^{A}\left(  P\right)  \text{
\ \ \ \ \ \ \ \ \ }A,B=1,2 \tag{59}%
\end{equation}
where, due to the Ambrose-Singer theorem [16], the key link of the theory is
given by%
\begin{align}
U_{A}^{B}\left(  P\right)   &  =\delta_{A}^{B}+\mathcal{R}_{A\mu\nu}%
^{B}dx^{\mu}\wedge dx^{\nu}\tag{60}\\
&  =\delta_{A}^{B}+\omega^{k}\left(  \mathcal{T}_{k}\right)  _{A}^{B}\nonumber
\end{align}

then%
\begin{equation}
\mathcal{R}_{A\mu\nu}^{B}dx^{\mu}\wedge dx^{\nu}\equiv\omega^{k}\left(
\mathcal{T}_{k}\right)  _{A}^{B} \tag{61}%
\end{equation}
immediately we can make the folllowing observations:

i) there exists a true and direct correspondence Manifold group structure,
tangent space, curvature and physical fields.

ii) the reason of the interplay described in i) is due to the unified
character of the theory: all the "matter and energy" content come from the
same spacetime manifold.

iii) the underlying (super) symmetry is quite evident from the link given
above: the curvature involves fermionic and bosonic structues (e.g. mixed
indices), then is not difficult to see that other fields with different amount
of spin can appear. Even more, due to the geometrical and group theoretical
meaning of the above expression, the possible transformations have local
(diffeomorphyc) character that make the role of the supersymmetry and the role
of the supergravity and superspace concept to be taken under consideration.

\section{Incompleteness of Rothstein's theorems: physics geometrization vs.
supermanifold construction}

\subsection{Poisson structure, quantization and supersymmetry}

Symplectic geometry grew out of the theoretical study of classical and quantum
mechanics. At first it was thought that it differs considerably from
Riemannian geometry, which developed from the study of curves and surfaces in
three dimensional Euclidean space, and went on to provide the language in
which General Relativity is studied. This fact was understandable given that
symplectic geometry started from the study of phase spaces for mechanical
systems but, with the subsequent seminal works of Cartan that introduce the
symplectic structure into the geometry of the spacetime calculus, that
thinking changed radically.

The existence of a symplectic structure on a manifold is a very significant
constraint and many simple and natural constructions in symplectic geometry
lead to manifolds which cannot possess a symplectic structure (or to spaces
which cannot possess a manifold structure). However these spaces often inherit
a bracket of functions from the Poisson bracket on the original symplectic
manifold. It is a (semi-)classical limit of quantum theory and also is the
theory dual to Lie algebra theory and, more generally, to Lie algebroid theory.

Poisson structures are the first stage in quantization, in the specific sense
that a Poisson bracket is the first term in the power series of a deformation
quantization. Poisson groups are also important in studies of complete integrability.

From the point of view of the Poisson structure associated to the differential
forms induced by the unitary transformation from the G-valuated tangent space
implies automatically, the existence of an \textit{even non-degenerate
(super)metric. }The remaining question of the previous section was if the
induced structure from the tangent space (via Ambrose-Singer theorem) was
intrinsically related to a supermanifold structure (e.g.hidden supersymmetry,
etc.). Some of these results were pointed out in the context of
supergeometrical analysis by Rothstein and by others authors [17,15],
corroborating this fact in some sense. Consequently we have actually several
models coming mainly from string theoretical frameworks that are potentially
ruled out. Let us see this issue with more detail: from the structure of the
tangent space $T_{p}\left(  M\right)  $ we have seen%
\begin{align}
U_{A}^{B}\left(  P\right)   &  =\delta_{A}^{B}+\mathcal{R}_{A\mu\nu}%
^{B}dx^{\mu}\wedge dx^{\nu}\tag{62}\\
&  =\delta_{A}^{B}+\omega^{k}\left(  \mathcal{T}_{k}\right)  _{A}^{B}\nonumber
\end{align}
where the Poisson structure is evident (as the dual of the Lie algebra of the
group manifold) in our case leading to the identification%
\begin{equation}
\mathcal{R}_{A\mu\nu}^{B}dx^{\mu}\wedge dx^{\nu}\equiv\omega^{k}\left(
\mathcal{T}_{k}\right)  _{A}^{B} \tag{63}%
\end{equation}
We have in the general case, a (matrix) automorphic structure. The general
translation to the spacetime from the above structure in the tangent space
takes the form%
\begin{align}
\widetilde{\omega}  &  =\frac{1}{2}\left[  \omega_{ij}+\frac{1}{2}\left(
\omega_{kl}\left(  \Gamma_{\text{ \ }ai}^{k}\Gamma_{\text{ \ }bj}^{l}%
-\Gamma_{\text{ \ }bj}^{k}\Gamma_{\text{ \ }ai}^{l}\right)  +g_{bd}R_{ija}%
^{d}\right)  d\psi^{a}d\psi^{b}\right]  dx^{i}\wedge dx^{j}+\omega
_{ij}A_{\text{ \ }bm}^{j}dx^{m}dx^{i}d\psi^{b}+\tag{64}\\
&  +\frac{1}{2}\left[  g_{ab}+\frac{1}{2}\left(  g_{cd}\left(  \Gamma_{\text{
\ }ib}^{c}\Gamma_{\text{ \ }ja}^{d}-\Gamma_{\text{ \ }ja}^{c}\Gamma_{\text{
\ }ib}^{d}\right)  +\omega_{lj}R_{abi}^{l}\right)  dx^{i}\wedge dx^{j}\right]
d\psi^{a}d\psi^{b}+g_{ab}A_{\text{ \ }id}^{b}d\psi^{d}d\psi^{a}dx^{i}\nonumber
\end{align}
Because covariant derivatives are defined in the usual (group theoretical)
way
\begin{align}
D\psi^{a}  &  =d\psi^{a}-\Gamma_{\text{ \ }ib}^{i}d\psi^{b}dx^{i}\tag{65}\\
Dx^{i}  &  =dx^{i}-\Gamma_{\text{ }aj}^{i}\text{ }dx^{j}d\psi^{a} \tag{66}%
\end{align}
we can rewrite $\widetilde{\omega}$ in a compact form as%
\begin{equation}
\widetilde{\omega}=\frac{1}{2}\left[  \left(  \omega_{ij}Dx^{i}\wedge
Dx^{j}+\frac{1}{2}g_{bd}R_{ija}^{d}d\psi^{a}d\psi^{b}dx^{i}\wedge
dx^{j}\right)  +\left(  g_{ab}D\theta^{a}D\theta^{b}+\frac{1}{2}\omega
_{lj}R_{abi}^{l}dx^{i}\wedge dx^{j}d\theta^{a}d\theta^{b}\right)  \right]
\tag{67}%
\end{equation}
At the tangent space, where that unitary transformation makes the link, the
first derivatives of the metric are zero, remaining only the curvatures, we
arrive to%
\begin{equation}
\widetilde{\omega}=\frac{1}{2}\left[  \left(  \eta_{ij}+\frac{1}{2}%
\epsilon_{bd}R_{ija}^{d}d\psi^{a}d\psi^{b}\right)  dx^{i}\wedge dx^{j}+\left(
\epsilon_{ab}+\frac{1}{2}\eta_{lj}R_{abi}^{l}dx^{i}\wedge dx^{j}\right)
d\psi^{a}d\psi^{b}\right]  \tag{68}%
\end{equation}

Here the Poisson structure can be checked%
\begin{align}
\eta_{ij}+\frac{1}{2}\epsilon_{bd}R_{ija}^{d}d\psi^{a}d\psi^{b}  &  =\left(
\delta_{j}^{k}+\frac{1}{2}\epsilon_{bd}\eta^{kl}R_{lja}^{d}d\psi^{a}d\psi
^{b}\right)  \eta_{ki}\tag{69}\\
\epsilon_{ab}+\frac{1}{2}\eta_{lj}R_{abi}^{l}dx^{i}\wedge dx^{j}  &  =\left(
\delta_{b}^{c}+\frac{1}{2}\eta_{lj}\epsilon^{cd}R_{dbi}^{l}dx^{i}\wedge
dx^{j}\right)  \epsilon_{ac} \tag{70}%
\end{align}
In expressions (64-70)\ the curvatures, the differential forms and the other
geometrical operators depend also on the field where they are defined:
$\mathbb{R}$, $\mathbb{C}$\ or $\mathbb{H}$. In the quaternionic $\mathbb{H}%
$-case (that can correspond to the SU(2)-structure of the UFT\ of Borchsenius
for example) the metric is quaternion valuated with the propierty
$\omega_{\left[  ij\right]  }^{\dagger}=-\omega_{\left[  ji\right]  }$ and the
covariant derivative can be straightforwardly defined as expressions
(65,66)\ but with the connection and coordinates also quaternion valuated. The
fundamental point in a such a case going towards a fully reliable
gravitational theory is to fix the connection in order to have a true link
with the physical situation. The matrix representation of structures (69,70)
are automorphic ones: e.g. they belong to the identity and to the symplectic
block generating the corresponding trascendent (parameter depending)
functions. Now, we will analize the above fundamental structure under the
light of the supersymplectic structure given by Rothstein (notation as in Ref.
[17])%
\begin{equation}
\widetilde{\omega}=\frac{1}{2}\left(  \omega_{ij}+\frac{1}{2}g_{bd}R_{ija}%
^{d}\theta^{a}\theta^{b}\right)  dx^{i}dx^{j}+g_{ab}D\theta^{a}D\theta^{b}
\tag{71}%
\end{equation}
where the usual set of \ Grassmann supercoordinates were introduced:
$x^{1},....x^{j};\theta^{1}.....\theta^{d};$ the superspace metrics were
defined as: $\omega_{ij}=\left(  \frac{\partial}{\partial x^{i}}%
,\frac{\partial}{\partial x^{j}}\right)  ,g_{ab}=\left(  \frac{\partial
}{\partial\theta^{a}},\frac{\partial}{\partial\theta^{b}}\right)  $ and%
\begin{equation}
\nabla_{\frac{\partial}{\partial x^{i}}}\left(  \theta^{a}\right)  =A_{\text{
\ }ib}^{i}\theta^{b} \tag{72}%
\end{equation}

Due to the last expression, we can put $\widetilde{\omega}$ in a compact form
with the introduction of a suitable covariant derivative: $D\theta^{a}%
=d\theta^{a}-A_{\text{ \ }ib}^{i}\theta^{b}dx^{i}.$ With all the definitions
at hands, the Poisson structure of $\widetilde{\omega}$ in the case of
Rothstein's is easily verified%
\begin{equation}
\omega_{ij}+\frac{1}{2}g_{bd}R_{ija}^{d}\theta^{a}\theta^{b}=\left(
\delta_{i}^{k}+\underset{\equiv B}{\underbrace{\frac{1}{2}g_{bd}\omega
^{lk}R_{ila}^{d}\theta^{a}\theta^{b}}}\right)  \omega_{kj} \tag{73}%
\end{equation}
The important remark of Rothstein [17] is that the matrix representation of
the structure$B$ has \textit{nilpotent} entries, schematically%
\begin{equation}
\widetilde{\omega}^{-1}=\left[  \omega^{-1}\left(  I-B+B^{2}-B^{3}....\right)
\right]  ^{ij}\nabla_{i}\wedge\nabla_{j}+g^{ab}\frac{\partial}{\partial
\theta^{a}}\wedge\frac{\partial}{\partial\theta^{b}} \tag{74}%
\end{equation}
where, as is obvious $B^{n}=0$ for $n>1$ and $n\in\mathbb{N}$

\emph{Remarks:}

from the above analysis, we can compare the Rothstein case with the general
one arriving to the following points:

i) In the Rothstein case only a part of the full induced metric from the
tangent space is preserved ("one way" extension [11-14,17])

ii) The geometrical structures (particularly, the fermionic ones) are extended
"by hand" motivated, in general, to give by differentiation of the
corresponding closed forms, the standard supersymmetric spaces (e.g. Kahler,
$CP^{n}$, etc.) [17]. In fact it is easily seen from the structure of the
covariant derivatives: in the Rothstein case there are Grassmann coordinates
instead of the coordinate differential 1-forms contracted with the connection.

iii) In the Rothstein case the matrix representation (73) coming from the
Poisson structure is nilpotent (characteristic of Grassmann manifolds) in
sharp contrast with the general representation (68-70) coming from the tangent
space of the UFT\ that is automorphic.

\begin{remark}
was noted in [13] that the following facts arise:\emph{i)} A Grassmann
algebra, as used in supersymmetry, is equivalent, in some sense, to the spin
representation of a Clifford algebra. \emph{ii)} The questions about the
nature and origin of the vector space on which this orthogonal group acts are
completely open. \emph{iii)} If it is a tangent space or the space of a local
internal symmetry, the vectors will be functions of space-time, and the
Clifford algebra will be local. \emph{iv)} In other cases we will have a
global Clifford algebra. Consequently, the geometric structure of the UFT
presented here falls precisely in such a case.
\end{remark}

\subsection{UFT\ and supermanifold structure}

The\ UFT\ structure induced from the tangent space by means of the
Ambrose-Singer [16] theorem (62,63) verifies straigforwardly the
Darboux-Kostant theorem: e.g. it has a supermanifold structure.
Darboux-Kostant's theorem [15] is the supersymmetric generalization of
Darboux's theorem and statement that:

Given a $(2n|q)$-dimensional supersymplectic supermanifold $(M,\mathcal{A}%
_{M},\omega)$, it states that for any open neighbourhood $U$ of some point $m$
in $M$ there exists a set $(q_{1},...,q_{n},p1,...,pn;\xi_{1},...,\xi_{q})$ of
local coordinates on $V\mathcal{E}(U)$ so that $\omega$ on $U$ can be written
in the following

form,%
\begin{equation}
\left.  \omega\right\vert _{U}\equiv\widetilde{\omega}=\underset{i=1}%
{\overset{n}{\sum}}dpi\wedge dq^{i}+\underset{a=1}{\overset{q}{\sum}}%
\frac{\epsilon}{2}\left(  \xi^{a}\right)  ^{2}\text{ ,\ \ \ \ \ \ \ \ \ }%
\left(  \epsilon=\pm1\right)  \text{\ \ \ \ \ } \tag{75}%
\end{equation}

.

\begin{proof}
by simple inspection we can easily see that the expression (68) has the
structure (75). That means that we have locally a supersymplectic vector
superspace induced (globally) by a supersymplectic supermanifold.
\end{proof}

\section{Concluding discussion and perspectives}

Here we discuss some of the results obtained in this work and describe their
possible generalizations. We also briefly state other results as follows

From the point of view of the geometry and unification:

\begin{itemize}
\item i) The cornerstone of a consistent UFT must be a $G$-structure (for the
tangent bundle $T(M)$) which reflects the symmetries of the different fields considered.

\item ii) The difference between the QFT here and the standard QFT in curved
spacetime is that whilst the latter does not alter the spacetime structure
(whose structure group remains Lorentzian), the former alters the spacetime
structure radically since the structure group for the (reduced) tangent bundle
is now the correspondent to the induced QFT (the same curvature of the tangent space)

\item iii) The radical difference between spacetime signature and
non-hermitian dynamic operators is induced by the same G-structure.

\item iv) Torsion, through its dual four-dimensional vector, plays a key role
both in the signature of spacetime and the CP invariant character of the field dynamics.

\item v) From points iii) and iv) is clear that fermionic phase transitions in
the early universe as the paradigm of energy and dark matter could have a
satisfactory explanation seriously considering a theory as presented here
endowed with a G structure.
\end{itemize}

From the point of view of the boson-fermion symmetries

\begin{itemize}
\item iv) the Darboux-Kostant theorem is fulfilled in our case showing that M
fits the characteristic of a general supermanifold in addition to all those
the considerations given in [13,15,17].

\item v) The Rothstein theorem is incomplete to decribe the spacetime manifold
being it with a more general structure from the algebraic and geometrical viewpoint.
\end{itemize}

\textbf{Outlook}: there are several toipics that must be analyzed in future works:

\begin{itemize}
\item vi) There exists a deep relation of our research with early works where
quaternionic and even octonionic structures (as the Moffat-Boer theory) were
considered in the context of gravity: will be good to make a deep study of
this issue considering the boson-fermion symmetry and the link with the
quantum-gravity trouble.

\item vii) the possibility, following an old Dirac's conjecture, to find a
discrete quaternionic structure inside the Poincare group: this fact will be
give us the possibility of spacetime discretization without break Lorentz symmetries.

\item viii)The introduction of group theoretical methods of compactification
as in [28]

\item ix) the relation with nonlinearly realized symmetries and quantization.
\end{itemize}

\section{Acknowlegments}

I am very grateful to the JINR Directorate and the BLTP for his hospitality
and finnancial support. This work is devoted tothe memory of the Prof.
Academician Vladimir Georgievich Kadyshevsky that suddenly pass away this year .

\section{Appendix: Generalized Hodge-de Rham decomposition, the vector torsion
$h$ and the fermion interaction}

As pointed out in references[3-5,7,10] the torsion vector $h=h_{\alpha
}dx^{\alpha}$ (the 4-dimensional dual of the torsion field $T_{\beta
\gamma\delta}$) plays multiple roles and can be constrained in several
different physical situations. Mathematically, it is defined by the Hodge-de
Rham decomposition given by the \textbf{4-dimensional Helmholtz theorem} which states:

\textit{If $h=h_{\alpha}dx^{\alpha}$ $\notin F^{\prime}\left(  M\right)  $ is
a 1-form on $M$, then there exist a zero-form $\Omega$, a 2-form
$\alpha=A_{\left[  \mu\nu\right]  }dx^{\mu}\wedge dx^{\nu}$ and a harmonic
1-form $q=q_{\alpha}dx^{\alpha}$ on $M$ that}%
\begin{equation}
h=d\Omega+\delta\alpha+q\rightarrow h_{\alpha}=\nabla_{\alpha}\Omega
+\varepsilon_{\alpha}^{\beta\gamma\delta}\nabla_{\beta}A_{\gamma\delta
}+q_{\alpha}\,. \tag{76}%
\end{equation}
Notice that even if it is not harmonic, and assuming that $q_{\alpha}=$
$\left(  P_{\alpha}-eA_{\alpha}\right)  $ is a vector, an axial vector can be
added so that the above expression takes the form%
\begin{align}
h_{\alpha}  &  =\nabla_{\alpha}\Omega+\varepsilon_{\alpha}^{\beta\gamma\delta
}\nabla_{\beta}A_{\gamma\delta}+\varepsilon_{\alpha}^{\beta\gamma\delta
}M_{\beta\gamma\delta}+\left(  P_{\alpha}-eA_{\alpha}\right) \tag{77}\\
&  =\nabla_{\alpha}\Omega+\varepsilon_{\alpha}^{\beta\gamma\delta}%
\nabla_{\beta}A_{\gamma\delta}+\gamma^{5}b_{\alpha}+\left(  P_{\alpha
}-eA_{\alpha}\right)  \,, \tag{78}%
\end{align}
where $M_{\beta\gamma\delta}$ is a completely antisymmetric tensor. In such a
way, $\varepsilon_{\alpha}^{\beta\gamma\delta}M_{\beta\gamma\delta}$
$\equiv\gamma^{5}b_{\alpha}$ is an axial vector.

One can immediately see that, due to the theorem given above, one of the roles
of $h_{\alpha}$ is precisely to be a generalized energy-momentum vector,
avoiding the addition "by hand" of a matter Lagrangian in the action. As it is
well known, the addition of the matter Lagrangian leads, in general, to
non-minimally coupled terms into the equations of motion of the physical
fields. Consequently, avoiding the addition of energy-momentum tensor, the
fields and their interactions are effectively restricted thanks to the same
geometrical structure in the space-time itself.

\section{References}

[1] H. Weyl, \textquotedblleft Space-Time-Matter\textquotedblright, Dover (1952).

[2] Yu Xin, 1996, \textquotedblleft General Relativity on Spinor-Tensor
Manifold\textquotedblright, in: \textquotedblleft Quantum Gravity - Int.School
on Cosmology \& Gravitation\textquotedblright, XIV Course. Eds. P.G. Bergman,
V. de.Sabbata \& H.J. Treder, pp. 382-411, World Scientific.

[3] D.J. Cirilo-Lombardo, Int.J.Theor.Phys. \textbf{49}, 1288, (2010).

[4] D.J. Cirilo-Lombardo, Int.J.Theor.Phys. \textbf{50}, 1699 (2011).

[5] D.J. Cirilo-Lombardo, Int.J.Theor.Phys. \textbf{50}, 3621 (2011).

[6] D.J. Cirilo-Lombardo, J.Math.Phys.\textbf{\ 48}, 032301, (2007);
Class.Quant.Grav. \textbf{22 }, 4987 (2005).

[7] D.J Cirilo-Lombardo, Astropart.Phys. \textbf{50-52}, 51 (2013).

[8] V.B. Beresteskii, E.M. Lifshitz and L.P. Pitaevskii, \textit{Quantum
electrodynamics}, Pergamon Press, New York (1982).

[9] E. Majorana, \textquotedblleft Teoria Simmetrica Dell' Elettrone E Del
Positrone,\textquotedblright\ Il Nuovo Cimento (1924-1942), Vol. \textbf{14},
No. 4, pp. 171-184 (1937).

[10] D.J Cirilo-Lombardo, Physics of Particles and Nuclei, Vol. \textbf{44},
No. 5, pp.848--865 (2013)

[11] M. Pavsic, Adv. Appl. Cliff ord Algebras\textbf{ 22}, 449--481, (2012)

[12] Albert, A. A.: Structure of Algebras, Amer. Math. Soc., Providence, R.I., (1961).

[13] J.O. Winnberg, J. Math. Phys. \textbf{18} 625, (1977), M. Pavsic
J.Phys.Conf.Ser. \textbf{33} 422-427 (2006).

[14]\ N. A. Salingaros and G. P. Wene, Acta Applicandae Mathematicae
\textbf{4}, 27 1-292.(1985); M. Pavsic Adv.Appl.Clifford Algebras\textbf{ 20},
781-801 (2010) , Phys.Lett. \textbf{B692,} 212-217 (2010)

[15] Kostant, B., in: Lecture Notes in Mathematics vol\textbf{.570} , 177,
(Bleuler, K. and Reetz, A. eds), Proc. Conf. on Diff. Geom. Meth. in Math.
Phys., Bonn 1975., Springer-Verlag, Berlin, 1977.

[16] W. Ambrose and I. M. Singer, Transactions of the American Mathematical
Society, Vol. \textbf{75}, No. 3, pp. 428-443 (Nov., 1953).

[17] Rothstein, M., in: Lecture Notes in Physics vol.\textbf{375 }, 331
(Bartocci, C., Bruzzo, U.,and Cianci, R., eds), Proc. Conf. on Diff. Geom.
Meth. in Math. Phys., Rapallo

1990., Springer-Verlag, Berlin, 1991; C. Bartocci, U. Bruzzo and D. Hernandez
Ruiperez, \textit{The geometry of supermanifolds}, Kluwer, Dordrecht,

The Netherlands 1991

[18] I. Bars and S. H. Chen, Phys. Rev. \textbf{D 79}, 085021(2009).

[19] E. Witten, Phys. Rev.\textbf{ D 44} 314, (1991) 314.

[20] H. Ooguri and C. Vafa, Nucl. Phys. \textbf{B367}, 83 (1991); Nucl. Phys.
\textbf{B361, }469, (1991)

[21] E. Sezgin, \textit{Is there a stringy description of selfdual
supergravity in (2+2)-dimensions?}, Published in \textquotedblleft Trieste
1995, High energy physics and cos-

mology\textquotedblright\ 360-369; hep-th/9602099.

[22] Z. Khviengia, H. Lu, C.N. Pope, E. Sezgin, X.J. Wang and K.W. Xu, Nucl.
Phys. \textbf{B444,}468 (1995); hep-th/9504121.

[23] S. V. Ketov, Class. Quantum Grav. \textbf{10}, 1689, (1993); hep-th/9302091.

[24] S. V. Ketov, H. Nishino and S.J. Gates Jr., Phys. Lett.\textbf{ B 307, }
323, (1993); hep-th/9203081.

[25] M. A. De Andrade, O. M. Del Cima and L. P. Colatto, Phys. Lett. \textbf{B
370} 59 (1996); hep-th/9506146.

[26] M. F. Atiyah. and R.S. Ward, Commun. Math. Phys. \textbf{55, }117 (1977)

[27] C. M. Hull, JHEP \textbf{9811, }017 (1998); hep-th/9807127.

[28] D. J. Cirilo-Lombardo, Eur.Phys.J. \textbf{C72}, 2079, (2012)

[30] M. Pavsic, "The Landscape of theoretical physics: A Global view. From
point particles to the brane world and beyond, in search of a unifying
principle", (Kluwer 2001). 386 pp. (Fundamental theories of physics. 119).
e-Print: gr-qc/0610061

[31] R.L. Bishop and S.I. Goldberg, "Tensor analysis on manifolds" (Dover
1980) see in particular Chapter 5 pp.208 about the meaning of Lorentz metric

\end{document}